\documentclass[conference]{IEEEtran}

\usepackage{cite}
\usepackage{url}
\usepackage{amsmath,amssymb}
\usepackage{graphicx}
\usepackage{booktabs}
\usepackage{placeins}

\begin{document}

\title{Modeling Preferential Store Openings and Closings with Additively Consistent Industry Interactions}

\author{
\IEEEauthorblockN{Koutarou Tamura$^{*\dagger}$}
\IEEEauthorblockA{$^{*}$ Financial Technology Solution Division, Nomura Research Institute, Ltd., Tokyo, Japan\\
$^{\dagger}$ Institute of Systems and Information Engineering, University of Tsukuba, Tsukuba, Japan\\
k9-tamura@nri.co.jp, k.tamura.phd@gmail.com}
}

\maketitle

\begin{abstract}
Commercial agglomerations are formed by the openings and closings of individual stores, yet how these events shape the spatial distribution of stores remains unclear. Using geolocated store opening and closing events in Tokyo and Osaka, we directly estimate preferential location choice on square grids of several cell sizes. We propose a stochastic growth model that combines preferential selection of existing agglomerations, entry into empty cells, and inter-industry interactions. The model is additively consistent: the industry-specific location probabilities sum exactly to the overall location probability, so that inter-industry co-location can be modeled without altering the aggregate model. Generative simulations with the estimated parameters reproduce the heavy tail of the distribution of stores per cell; at 100\,m in Tokyo, the tail slope is $-1.81$ in the model versus $-1.82$ in the data. At larger cell sizes, generation from empty cells underestimates the number of occupied cells and yields a flatter tail. Inter-industry interactions improve the agreement of inter-industry correlations with the observations (100\,m) from 0.63 to 0.83 in Tokyo and from 0.40 to 0.87 in Osaka.
\end{abstract}


\section{Introduction}

Commercial agglomerations are closely tied to urban convenience, employment, and land prices, and the hollowing-out of city centers has become a problem in many cities~\cite{kickert2020}. Understanding where stores cluster and where they withdraw helps explain how cities form and change.

Large-scale statistical studies of economic agents have linked the rules of birth, growth, and death of individual agents to aggregate heavy-tailed distributions, mainly for firms and cities~\cite{gabaix1999,axtell2001,tamura2012,tamura2018,tamura_startup}; in inter-firm transaction networks, for example, preferential attachment has been observed and modeled~\cite{miura2012,ozaki2024}. For commercial agglomerations within cities, however, analyses have centered on the stock of stores or establishments at a single point in time, as in the scaling of urban hierarchies and industry composition~\cite{youn2016,bettencourt2013,mori2020}. How the stock relates to the flow---where new stores open and where closures occur---has rarely been tested directly, because official statistics are aggregated only every few years and cannot track individual openings and closings.

Store opening and closing information posted on the web captures commercial turnover at the event level, daily, and with geolocation~\cite{dataset,hfdataset}. It allows us to estimate directly where new stores open and where closures occur as a function of the existing number of stores, in the same spirit as measuring preferential attachment in growing networks~\cite{jeong2003,newman2001}. Because statistics aggregated over spatial units depend on the unit size (the modifiable areal unit problem, MAUP)~\cite{openshaw1984,arcaute2015}, we repeat the analysis at several cell sizes. Moreover, stores belong to industries, which exhibit same-industry clustering and competitive exclusion~\cite{eaton1979,jensen2006}. If an independent nonlinear location model is built for each industry, the sum of the industry-specific predictions does not agree with the aggregate model. We call the property that the industry-specific location probabilities sum exactly to the overall location probability \emph{additive consistency}, and we incorporate inter-industry interactions so that it holds. For practical analysis, the sum of industry-specific opening forecasts for a cell should agree with its aggregate opening forecast; otherwise, projected industry demand and overall commercial growth are inconsistent. Here we enforce this consistency within each cell.

Our contributions are threefold. (1) We reconstruct store opening and closing events in Tokyo and Osaka into operating spells and show, on grids of several cell sizes, that the distribution of stores per cell has a heavy tail and that industries co-locate in characteristic patterns (Section~\ref{sec:data}). (2) We propose an additively consistent stochastic model that combines preferential selection of existing agglomerations, entry into empty cells, and industry choice, and we derive the tail exponent under a mean-field approximation (Section~\ref{sec:model}). (3) We estimate the parameters from monthly events (Section~\ref{sec:estimation}) and show that generative simulations, given the estimates and the total number of stores at the validation date, reproduce the end-2024 distribution of stores per cell and the inter-industry co-location, which were not directly used for estimation (Section~\ref{sec:simulation}).

\section{Related Work}\label{sec:related}

\textbf{Retail location and agglomeration.} Retail location has been discussed in terms of clustering due to comparison shopping~\cite{eaton1979} and probabilistic trade-area models~\cite{huff1964}. Agglomeration has been measured by indices such as the Ellison--Glaeser index~\cite{ellison1997}, and Jensen estimated inter-industry attraction and repulsion coefficients from store proximity~\cite{jensen2006}. Other studies showed that the amenity mix of neighborhoods is predictable~\cite{hidalgo2020} and predicted new store locations from location-based services~\cite{karamshuk2013}. These studies mainly address the configuration of stores at a single point in time.

\textbf{Heavy tails in social phenomena.} Many social and economic quantities have heavy tails close to power laws, including city populations~\cite{gabaix1999}, firm sizes~\cite{axtell2001}, industry composition within cities~\cite{youn2016,bettencourt2013}, and urban hierarchies~\cite{mori2020}. Such estimates are known to depend on the definition of cities and the spatial units~\cite{openshaw1984,arcaute2015}.

\textbf{Preferential attachment as a source of power laws.} Simon's growth model~\cite{simon1955}, in which new elements join existing ones in proportion to their size and new elements are created at a constant rate, and the Barab\'asi--Albert model~\cite{barabasi1999} explain such heavy tails. With attachment probability proportional to $k^\alpha$, the degree distribution becomes a stretched exponential for $\alpha<1$ and condenses for $\alpha>1$~\cite{krapivsky2000}; initial attractiveness changes the tail exponent~\cite{dorogovtsev2000}, and node deletion changes the tail through the balance of addition and deletion~\cite{moore2006}. Preferential attachment has been measured in real data~\cite{jeong2003,newman2001,pham2015}, including business networks of firms, where it is a key ingredient of models that reproduce empirical laws~\cite{miura2012,ozaki2024}. Proportional growth (Gibrat's law) and urban growth models have also been studied extensively~\cite{sutton1997,makse1995,rozenfeld2008}. Unlike these studies, we estimate preference directly from individual opening and closing events and incorporate entry into new spatial units and industry choice. Preference in geographic space can be confounded with location-specific fitness~\cite{bianconi2001}; we regard the latter as absorbed into the number of stores. Occupancy is thus a proxy for location attractiveness, not a measure of causal reinforcement. Explicit covariates such as accessibility and population are needed to distinguish these effects.

\section{Dataset}\label{sec:data}

\subsection{Web-Posted Events and Operating Spells}\label{sec:dataset}
We use store openings and closings posted on Kaiten-Heiten.com~\cite{kaitenheiten}, distributed by Kobayashi et al.\ as a commercial dynamics dataset~\cite{dataset,hfdataset}. The data contain events in Tokyo and Osaka Prefectures from 2019 to 2025. After removing records with missing coordinates or dates, there are 30{,}463 events in Tokyo and 12{,}409 in Osaka. Each record has the event type, date, store name, address, industry (10 major categories), coordinates, and source URL. Food service and retail account for about 90\% of openings and 88\% of the stores operating at the end of 2024 (Table~\ref{tab:data}).

We reconstruct the events of each store into an operating spell that starts with an opening and ends with a closing; a store with only a closing (opening) observed is regarded as operating before (after) the observation period. Operating spells are reconstructed retrospectively using the full observation period. Most spells have only an opening (Tokyo 22{,}188, Osaka 9{,}628) or only a closing (Tokyo 7{,}617, Osaka 2{,}540) observed. The counts of events and reconstructed spells are summarized in Table~\ref{tab:spell_counts}. Repeated opening records were included in the event counts but merged during spell reconstruction (152 in Tokyo and 61 in Osaka); all closing events were used. Occupancy throughout this paper refers to reconstructed rather than census store counts.

\begin{table}[tb]
\centering
\caption{Events and stores by industry (2019--2025). Stores are those operating at the end of 2024.}\label{tab:data}
\setlength{\tabcolsep}{3pt}
\resizebox{\columnwidth}{!}{%
\begin{tabular}{lrrrrrr}
\toprule
 & \multicolumn{3}{c}{Tokyo} & \multicolumn{3}{c}{Osaka} \\
\cmidrule(lr){2-4}\cmidrule(lr){5-7}
Industry & Open & Close & Stores & Open & Close & Stores \\
\midrule
Food service & 15{,}431 & 4{,}030 & 13{,}634 & 6{,}443 & 1{,}008 & 5{,}569 \\
Retail & 4{,}618 & 2{,}507 & 4{,}498 & 2{,}149 & 1{,}012 & 2{,}082 \\
Auto/Real estate & 538 & 128 & 495 & 284 & 47 & 252 \\
Tourism/Lodging/Leisure & 484 & 467 & 472 & 255 & 236 & 251 \\
Beauty/Health/Welfare & 457 & 224 & 457 & 213 & 138 & 201 \\
Services & 473 & 222 & 442 & 210 & 70 & 199 \\
Sports & 301 & 89 & 272 & 119 & 33 & 110 \\
Rental & 159 & 137 & 149 & 46 & 49 & 45 \\
Public/Transport/Finance & 105 & 54 & 94 & 46 & 35 & 44 \\
Unclassified & 27 & 12 & 23 & 14 & 2 & 11 \\
\midrule
Total & 22{,}593 & 7{,}870 & 20{,}536 & 9{,}779 & 2{,}630 & 8{,}764 \\
\bottomrule
\end{tabular}}
\end{table}

\begin{table}[tb]
\centering
\caption{Event counts and reconstructed operating spells by city.}\label{tab:spell_counts}
{\footnotesize
\begin{tabular}{lrr}
\toprule
Category & Tokyo & Osaka \\
\midrule
\multicolumn{3}{l}{\textit{Observed events}} \\
\quad Opening events & 22{,}593 & 9{,}779 \\
\quad Closing events & 7{,}870 & 2{,}630 \\
\quad Total events & 30{,}463 & 12{,}409 \\
\quad Repeated openings merged$^{a}$ & 152 & 61 \\
\midrule
\multicolumn{3}{l}{\textit{Reconstructed operating spells}} \\
\quad Opening only & 22{,}188 & 9{,}628 \\
\quad Closing only & 7{,}617 & 2{,}540 \\
\quad Opening and closing & 253 & 90 \\
\quad Total spells & 30{,}058 & 12{,}258 \\

\midrule
\multicolumn{3}{l}{\textit{Stores operating}} \\
\quad Beginning of 2020 & 10{,}823 & 3{,}887 \\
\quad End of 2024 & 20{,}536 & 8{,}764 \\
\bottomrule
\multicolumn{3}{l}{\scriptsize $^{a}$ Included in opening events; not additional spells.} \\
\end{tabular}}
\end{table}

\subsection{Grid Cells, Occupancy, and Event Trends}\label{sec:basic}

\textbf{Grid cells and occupancy.} We divide the study area, excluding remote islands, into square grid cells of side length $\ell\in\{100,200,300,400,500,750,1000\}$\,m and use cells whose land-area fraction is at least 0.5, including empty cells. Let $S_{ij}^{(\ell)}(t)$ be the number of stores of industry $j$ in cell $i$ at time $t$; we call $s_i^{(\ell)}(t)=\sum_jS_{ij}^{(\ell)}(t)$ the \emph{occupancy} of the cell. We use a monthly panel from January 2020 to December 2024 (stores at the beginning of each month and openings and closings within it), constructed so that $s_i(t+1)=s_i(t)+O_i(t)-C_i(t)$ holds exactly for every cell, where $O_i(t)$ and $C_i(t)$ are the monthly opening and closing counts. Panel openings count reconstructed spell starts rather than repeated opening records.

\begin{figure*}[tb]
\centering

\includegraphics[width=0.97\textwidth]{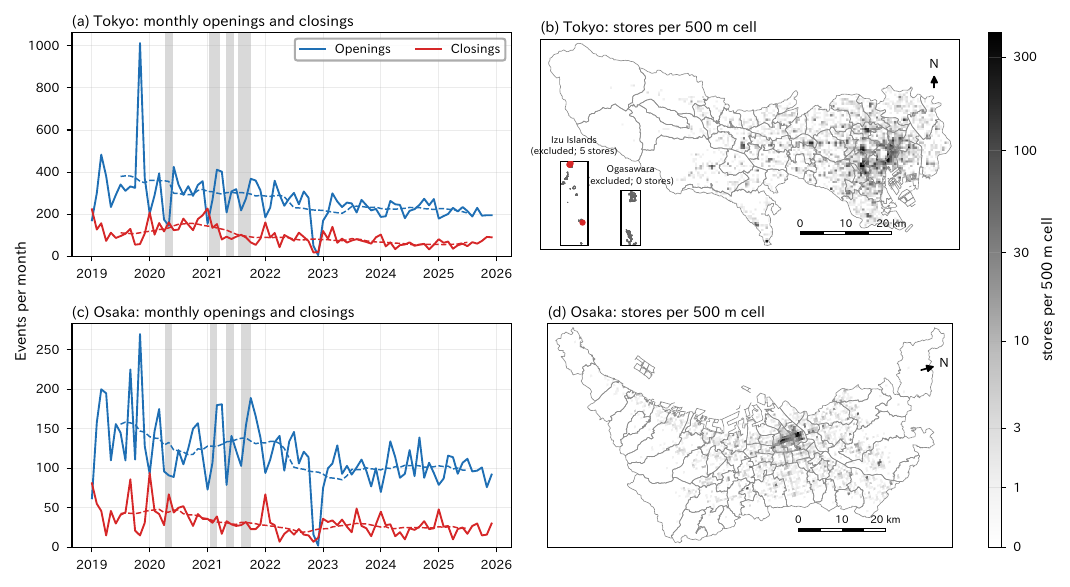}
\caption{Monthly events and occupancy. (a)(c) Monthly openings and closings (dashed: 12-month moving averages; shading: states of emergency). (b)(d) Occupancy $s$ of 500\,m cells at the end of 2024 (common color scale; empty cells in white; the Osaka map is rotated, arrow pointing north).}
\label{fig:data}
\end{figure*}

\textbf{Event trends.} Fig.~\ref{fig:data} shows occupancy maps and monthly events. In the estimation period (2020--2023), the 100\,m cells contain 12{,}688 openings and 5{,}064 closings in Tokyo and 5{,}462 openings and 1{,}526 closings in Osaka. The closing-to-opening ratio is $r=0.399$ in Tokyo and $r=0.279$ in Osaka. Osaka has almost the same area as Tokyo but only about 40\% as many stores.

\subsection{Occupancy Distribution and Industry Co-location}\label{sec:empirical}

\begin{figure}[tb]
\centering

\includegraphics[width=\linewidth]{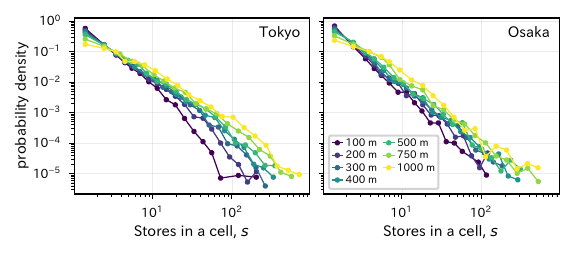}
\caption{Occupancy distribution (end of 2024, occupied cells, log-binned density). Left: Tokyo; right: Osaka. Colors indicate the cell size.}
\label{fig:sdist}
\end{figure}

\textbf{Heavy-tailed occupancy.} For both cities and all cell sizes, the distribution of occupancy $s$ over occupied cells has a heavy tail that decays approximately linearly on log--log axes (Fig.~\ref{fig:sdist}). As $\ell$ increases, the distribution shifts toward larger $s$ and its tail becomes flatter; the tail exponent thus varies with the unit of analysis (Table~\ref{tab:comparison}). The maximum occupancy of a 100\,m cell (229 stores in Tokyo, 129 in Osaka) reflects many stores in large shopping facilities recorded at the same coordinates.

\textbf{Co-location ratio.} We measure which industries locate together by counting pairs of stores within cells. The number of store pairs of industries $j$ and $k$, summed over cells, is divided by its expectation when industry labels are randomly permuted among stores (with the number of stores in each cell and in each industry fixed):
\begin{equation}
R_{jk}=\frac{\sum_iS_{ij}S_{ik}}{Q\,N_jN_k/\{N(N-1)\}}\ (j\ne k),\quad Q=\sum_is_i(s_i-1),
\label{eq:cooc}
\end{equation}
where $N_j$ is the number of stores in industry $j$; for $j=k$, $S_{ij}S_{ik}$ and $N_jN_k$ are replaced by $S_{ij}(S_{ij}-1)$ and $N_j(N_j-1)$. $R_{jk}>1$ ($<1$) means that $j$ and $k$ share cells more (less) often than by chance. Because cell sizes are fixed in the null model, $R$ excludes the effect that large cells contain many stores of every industry.

\begin{figure}[tb]
\centering
\includegraphics[width=\linewidth]{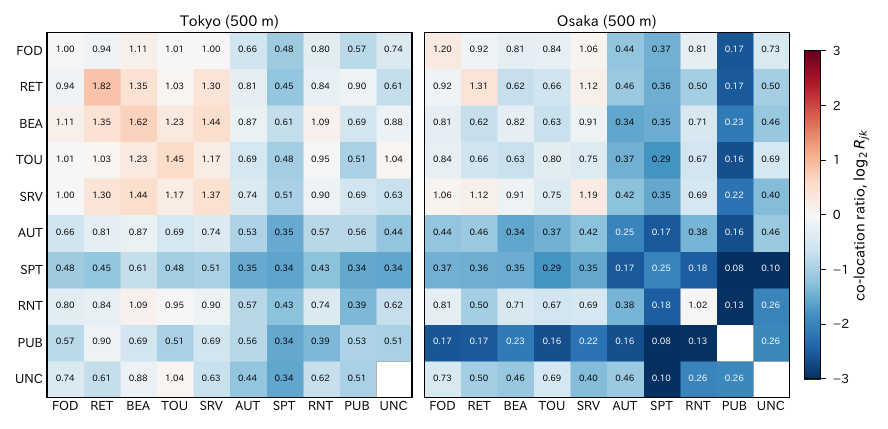}
\caption{Co-location ratio $R_{jk}$ (Eq.~(\ref{eq:cooc}), 500\,m, end of 2024; left: Tokyo, right: Osaka). Colors show $\log_2R_{jk}$; numbers are $R_{jk}$. FOD: food service; RET: retail; BEA: beauty/health/welfare; TOU: tourism/lodging/leisure; SRV: services; AUT: auto/real estate; SPT: sports; RNT: rental; PUB: public/transport/finance; UNC: unclassified.}
\label{fig:industry_stock}
\end{figure}

At 500\,m (Fig.~\ref{fig:industry_stock}), retail shows strong same-industry clustering (Tokyo 1.82, Osaka 1.31). Everyday shopping and personal services co-locate (e.g., services $\times$ beauty 1.44 in Tokyo), whereas food service and retail slightly avoid each other in both cities (0.94, 0.93). Sports, public/transport/finance, and auto/real estate have $R<1$ with most industries and locate away from commercial agglomerations. The patterns are broadly common to the two cities; beauty/health/welfare in Osaka clusters at 100\,m (1.33) but not at 500\,m (0.82). At 100\,m, retail clustering is strong (Tokyo 5.06, Osaka 2.76), reflecting shopping facilities.

\section{Proposed Model}\label{sec:model}
The heavy tail can arise from preferential selection~\cite{simon1955,barabasi1999}. We build a stochastic model consisting of (A) preferential selection of occupied cells for openings and closings, (B) entry into empty cells, and (C) industry choice, and in (D) we derive the tail exponent.

\subsection{Preferential Selection of Occupied Cells}\label{sec:kernel}
We regard each opening (closing) in month $t$ as a multinomial choice of a cell as a function of the occupancy $s_i(t)$ at the beginning of the month. Among occupied cells, an opening and a closing occur in cell $i$ with probabilities
\begin{align}
P_+(i)&=\frac{(s_i(t)+c_+)^{\alpha_+}}{\sum_{l:\,s_l(t)>0}(s_l(t)+c_+)^{\alpha_+}}, \label{eq:kernelp}\\
P_-(i)&=\frac{(s_i(t)+c_-)^{\alpha_-}}{\sum_{l:\,s_l(t)>0}(s_l(t)+c_-)^{\alpha_-}}.
\label{eq:kernel}
\end{align}
The \emph{preference exponents} $\alpha_\pm$ indicate linear ($\alpha=1$, Barab\'asi--Albert type~\cite{barabasi1999}), sublinear ($\alpha<1$), or superlinear ($\alpha>1$) preference. The offsets $c_\pm>-1$ act as $c_\pm$ virtual stores. Positive (negative) offsets increase (decrease) the relative weights of small agglomerations compared with the zero-offset case.

\subsection{Empty-Cell Selection}\label{sec:entry}
Extending Eq.~(\ref{eq:kernelp}) to empty cells with weight $c^{\alpha_+}$ ($c>0$) would give them nearly constant total weight $c^{\alpha_+}M_0$ ($M_0$: number of empty cells), so most openings would enter empty cells at small $N$ and few at large $N$. We therefore introduce the \emph{empty-cell selection rate} $\pi$, the constant fraction of openings that go to empty cells, as a spatial version of Simon's model~\cite{simon1955}. An opening chooses cell $i$ with probability
\begin{equation}
p_i(t)=
\begin{cases}
\dfrac{\pi}{M_0(t)} & s_i(t)=0,\\[2ex]
(1-\pi)\,P_+(i) & s_i(t)>0.
\end{cases}
\label{eq:spatial}
\end{equation}
A closing chooses an occupied cell with $P_-(i)$ and closes one of its stores at random. Openings and closings occur at the ratio $1:r$.

\subsection{Industry Choice and Additive Consistency}\label{sec:industrymodel}
After cell $i$ is chosen, industry $j$ is chosen with probability
\begin{equation}
q_{ij}(t)=\frac{\exp\bigl\{\gamma_j+\sum_k\beta_{kj}\log(S_{ik}(t)+c_0)\bigr\}}{\sum_m\exp\bigl\{\gamma_m+\sum_k\beta_{km}\log(S_{ik}(t)+c_0)\bigr\}}
\label{eq:industry}
\end{equation}
($c_0=1$). Here, $\gamma_j$ is the baseline propensity of industry $j$, which determines the composition of first entrants into empty cells, and $\beta_{kj}$, the \emph{opening co-occurrence coefficient}, represents the association between existing industry $k$ and the choice of new industry $j$. Only differences $\beta_{kj}-\beta_{km}$ are identified, representing elasticities of the odds of $j$ relative to $m$ with respect to $S_{ik}+c_0$; we report $\beta$ centered so that $\sum_j\beta_{kj}=0$. The coefficient $\beta_{kj}$ is a conditional association (statistical attraction or repulsion), not a causal effect, and may include common factors such as land prices, building use, and distance to stations. Stores with a missing industry form an ``unclassified'' industry.

\begin{figure}[tb]
\centering
\includegraphics[width=\linewidth]{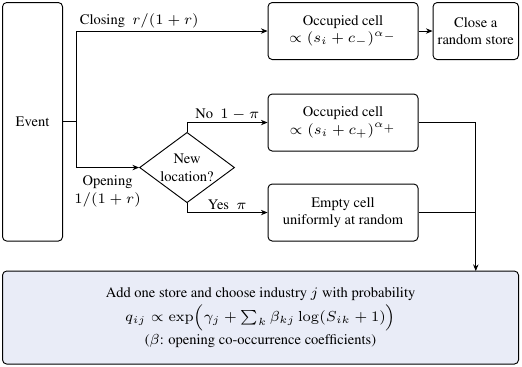}
\caption{Two-stage choice model. An opening chooses an empty cell (prob.\ $\pi$) or an occupied cell ($\propto(s+c_+)^{\alpha_+}$), then an industry by Eq.~(\ref{eq:industry}); a closing chooses a cell $\propto(s+c_-)^{\alpha_-}$ and a store at random.}
\label{fig:concept}
\end{figure}

\textbf{Additive consistency.} Summing the joint probability $P(i,j\mid t)=p_i(t)\,q_{ij}(t)$ over industries gives
\begin{equation}
\sum_jP(i,j\mid t)=p_i(t)\qquad(\forall\,i,t).
\label{eq:additivity}
\end{equation}
This identity itself follows from the chain rule of conditional probability; what matters is the design that makes it hold. Independent nonlinear models fitted to each industry generally do not sum to the aggregate model. In our model, the location rule $p_i$ is an industry-independent first stage, and inter-industry interactions are confined to the second-stage industry choice. Hence the occupancy dynamics are independent of industry classification and the value of $\beta$, and the generation of occupancy and that of industry composition can be estimated and validated separately (we use this when comparing models with and without $\beta$ in Section~\ref{sec:simulation}). The cost is that industry composition does not affect the choice of cells (Fig.~\ref{fig:concept}). This is a deliberate restriction to separate aggregate growth from industry composition; allowing feedback from composition to cell choice remains future work.

\subsection{Mean-Field Tail Exponent and Its Limitations}\label{sec:theory}
Under a mean-field approximation, we derive the tail exponent for the case in which existing agglomerations grow at a positive rate and the number of occupied cells increases in proportion to the total number of stores. Since $r$ closings occur per opening ($0\le r<1$), $N$ increases by $1-r$ per opening. With $W_\pm=\sum_{l:\,s_l>0}(s_l+c_\pm)^{\alpha_\pm}$, the expected change of a cell with occupancy $s$ is
\begin{equation}
\frac{ds}{dN}=\frac{1}{1-r}\Bigl[(1-\pi)\frac{(s+c_+)^{\alpha_+}}{W_+}-r\,\frac{(s+c_-)^{\alpha_-}}{W_-}\Bigr]\equiv\frac{g(s)}{N}.
\label{eq:dsdN}
\end{equation}
New cells are born in proportion to the increase in $N$. The net increase of occupied cells differs from this because single-store cells can be emptied by closings; we neglect such disappearance.

Let $\alpha_\pm=1$ and assume that $\mu=M^+/N$ ($M^+$: number of occupied cells) is constant. Then $W_\pm=N(1+c_\pm\mu)$, and for $s\gg|c_\pm|$,
\begin{equation}
\frac{ds}{dN}=\beta_N\frac{s}{N},\qquad
\beta_N=\frac{1}{1-r}\Bigl[\frac{1-\pi}{1+c_+\mu}-\frac{r}{1+c_-\mu}\Bigr],
\label{eq:betaN}
\end{equation}
so a cell born at $N_0$ grows as $s(N)\simeq(N/N_0)^{\beta_N}$. Since $N_0$ is uniform on $[0,N]$,
\begin{equation}
P(S\ge s)\simeq P\bigl(N_0\le N s^{-1/\beta_N}\bigr)=s^{-\tau},\qquad \tau=\frac{1}{\beta_N}
\label{eq:tau}
\end{equation}
($\beta_N>0$); without offsets, $\tau=(1-r)/(1-\pi-r)$. Larger $\pi$ or $r$, larger $c_+$, or smaller $c_-$ makes the tail lighter. Since $\mu$ is observed, $\tau$ needs no additional parameters. For $\alpha_\pm\ne1$, $P(S\ge s)=\exp\{-\int^s ds'/g(s')\}$ gives the local exponent
\begin{equation}
\tau(s)=-\frac{d\ln P(S\ge s)}{d\ln s}=\frac{s}{g(s)}.
\label{eq:taulocal}
\end{equation}
For $\alpha_+<1$, where the opening term dominates, $\tau(s)$ increases with $s$ and the tail steepens (cf.\ the stretched exponential without closings~\cite{krapivsky2000}).

This approximation ignores (a) the disappearance of single-store cells, (b) stochastic fluctuations of growth, (c) the finite number of empty cells, (d) spatial correlations between cells, and (e) the growth history before the observation period. (a) and (b) mainly affect small $s$, and (c) matters at large cell sizes with few empty cells. We therefore use Eq.~(\ref{eq:tau}) as a guide and compare with observations through simulations (Section~\ref{sec:simulation}).

If opening and closing events have capture rates $p_+$ and $p_-$, respectively, then $r_{\rm obs}\simeq(p_-/p_+)r_{\rm true}$. Missing closings can therefore change both the implied exponent and the growth conditions ($r<1$ and $\beta_N>0$). This does not invalidate reproduction of the reconstructed population, but limits generalization to all stores.

\textbf{Origin of the power-law tail.} The power-law tail arises from the spread of birth times $N_0$ combined with growth exponential in $\ln(N/N_0)$. With $\pi=0$, no new cells are born and birth times do not spread; when openings and closings balance, total occupancy does not grow. In either case, the derivation above no longer applies. The power-law mechanism thus relies on the continual supply of new agglomerations ($\pi>0$), growth in the number of stores ($r<1$), and positive growth of existing agglomerations ($\beta_N>0$) under the linear approximation.

\section{Parameter Estimation}\label{sec:estimation}
\begin{figure*}[tb]
\centering

\includegraphics[width=0.97\textwidth]{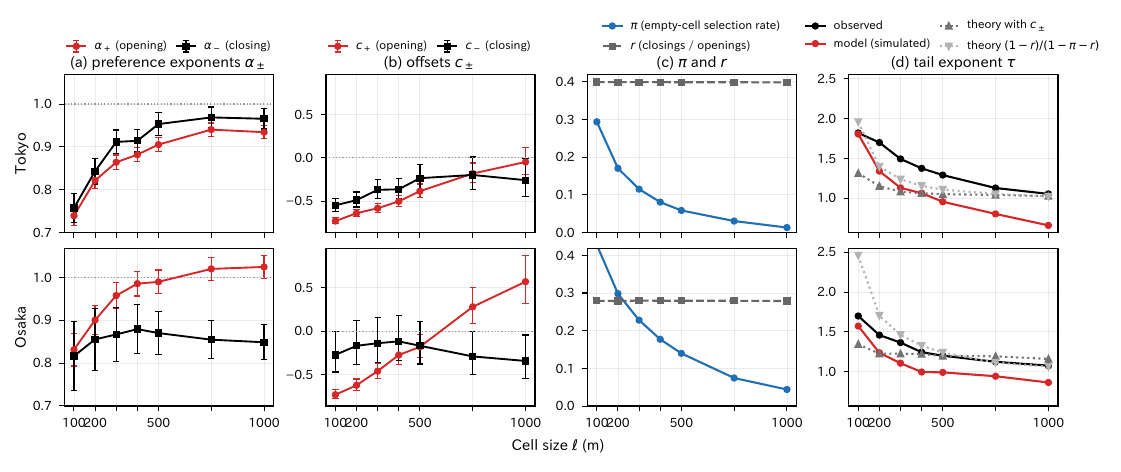}
\caption{Estimates versus cell size $\ell$ (top: Tokyo; bottom: Osaka). (a) Preference exponents $\alpha_\pm$ (95\% intervals). (b) Offsets $c_\pm$. (c) $\pi$ and $r$. (d) Tail exponent $\tau$: observed, simulated (Section~\ref{sec:simulation}), and Eq.~(\ref{eq:tau}) with and without offsets.}
\label{fig:alpha}
\end{figure*}

We estimate $\alpha_\pm$, $c_\pm$, $\pi$, $r$, $\gamma_j$, and $\beta_{kj}$ separately for each cell size from the 48 months of data from January 2020 to December 2023. The 2024 data are reserved for validation. Fig.~\ref{fig:alpha} shows the estimates.

\subsection{Preference Exponents and Offsets}\label{sec:est_kernel}
The parameters $\theta=(\alpha_\pm,c_\pm)$ are estimated by maximizing the event-level conditional logit likelihood
\begin{equation}
\log L(\theta)=\sum_{e}\Bigl[\log A(s_{i_e})-\log\sum_{l\in\mathcal{R}_e}A(s_l)\Bigr],
\label{eq:mle}
\end{equation}
where $A(s)=(s+c_\pm)^{\alpha_\pm}$, $e$ runs over events in the estimation period in cells occupied at the beginning of the corresponding month (openings in empty cells are summarized by $\pi$), $i_e$ is the cell of event $e$, and $\mathcal{R}_e$ is the set of cells occupied at the beginning of the month. We aggregate cells by occupancy in each month and obtain 95\% intervals from the numerical Hessian.

In both cities, $\alpha_\pm$ are sublinear for small cells ($\alpha_+=0.74$, $\alpha_-=0.76$ at 100\,m in Tokyo) and approach 1 as the cell size increases (Fig.~\ref{fig:alpha}(a)); only openings in Osaka at 750\,m and above are slightly superlinear (1.02). The offsets are negative for small cells ($c_+=-0.73$, $c_-=-0.55$ at 100\,m in Tokyo): the ratio of opening weights between $s=2$ and $s=1$ is about 3.1, well above 2 under proportional selection. Single-store cells thus attract relatively few openings.

\subsection{Empty-Cell Selection Rate and Closing-to-Opening Ratio}\label{sec:est_entry}
The rate $\pi$ is the fraction of openings in cells empty at the beginning of the month, and $r$ the ratio of closings to openings. At 100\,m, $\pi$ is 0.294 in Tokyo and 0.428 in Osaka, and $r$ is 0.399 and 0.279; the reconstructed store population is in a growth phase. The rate $\pi$ decreases by a factor of more than 20 from 100 to 1000\,m in Tokyo (Fig.~\ref{fig:alpha}(c)), because openings in new locations are absorbed within larger cells. The rate $\pi$ is larger in sparser Osaka at all cell sizes.

\subsection{Opening Co-occurrence Coefficients}\label{sec:est_industry}
The log-likelihood of $P(i,j\mid t)=p_i(t)q_{ij}(t)$ separates into cell-choice and industry-choice terms, so we estimate $(\gamma,\beta)$ by maximizing
\begin{equation}
\log L(\gamma,\beta)=\sum_e\log q_{i_ej_e}(t_e)-\lambda N_{\rm open}\sum_{k,j}\beta_{kj}^2,
\label{eq:mle_industry}
\end{equation}
where $e$ runs over openings, $j_e$ is the industry of the opening, and $q_{i_ej_e}(t_e)$ is computed from the stores by industry in cell $i_e$ at the beginning of the month. The coefficients of food service, the most frequent industry, are fixed to zero, and a weak L2 penalty ($\lambda=10^{-3}$; $N_{\rm open}$: number of openings) prevents divergence for rare industries. Compared with $\beta=0$, the per-event log-likelihood of the 2024 openings improves from $-0.905$ to $-0.854$ in Tokyo and from $-0.969$ to $-0.903$ in Osaka (100\,m). A same-industry-only model cannot explicitly represent cross-industry attraction or repulsion, although it can induce indirect correlations. We focus on the joint contribution of industry interactions; their separate contributions remain to be evaluated.

\begin{figure*}[tb]
\centering
\includegraphics[width=0.97\textwidth]{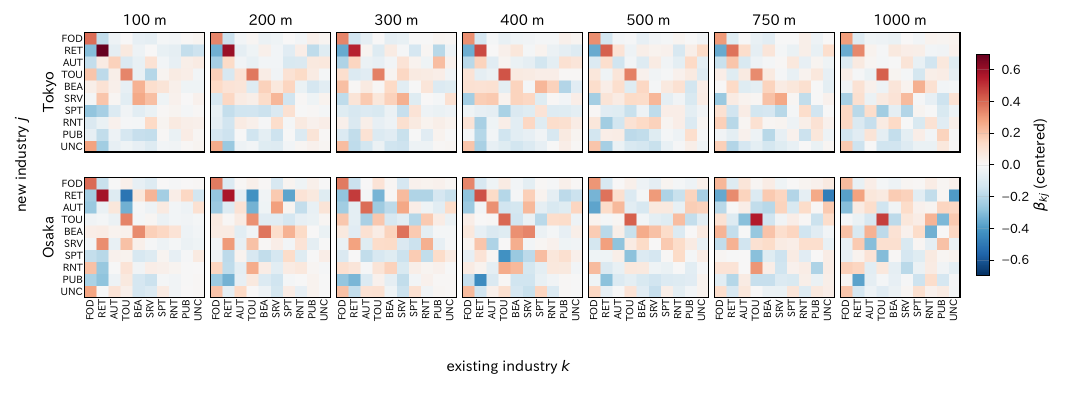}
\caption{Estimated $\beta_{kj}$ (centered; top: Tokyo, bottom: Osaka; 100--1000\,m from left). Rows: new industry $j$; columns: existing industry $k$; red: positive association; blue: negative. Abbreviations as in Fig.~\ref{fig:industry_stock}.}
\label{fig:beta}
\end{figure*}

Fig.~\ref{fig:beta} shows the estimated $\beta_{kj}$. In both cities and at all cell sizes, (i) the diagonal coefficients are positive for food service, retail, beauty/health/welfare, and tourism/lodging/leisure (same-industry clustering~\cite{eaton1979}); (ii) retail is less likely to open where food service is abundant, and vice versa; and (iii) unclassified stores are more likely to open where food service is abundant. The avoidance of retail-rich cells by food service weakens with cell size. Because food service, about two-thirds of openings, is the most probable industry in most cells, $\beta$ mainly improves the probabilities of the second and subsequent industries.

\subsection{Implied Tail Exponent and Robustness}\label{sec:est_summary}
The parameter that changes most with cell size is $\pi$, so the dependence on cell size, including that of the tail exponent through Eq.~(\ref{eq:tau}), arises mainly from how entry into new locations appears as entry into empty cells. At 100\,m in Tokyo, Eq.~(\ref{eq:tau}) gives $\tau_{\rm th}=1.31$ with offsets and 1.96 without them, bracketing the observed $\tau_{\rm obs}=1.82$ (Fig.~\ref{fig:alpha}(d)). As the cell size increases, $\tau_{\rm th}$ approaches 1 through the decrease in $\pi$, consistent with the decrease of $\tau_{\rm obs}$. The value with offsets is too small presumably because $\alpha_+<1$ increases the local exponent (Eq.~(\ref{eq:taulocal})).

\begin{figure*}[tb]
\centering
\includegraphics[width=0.94\textwidth]{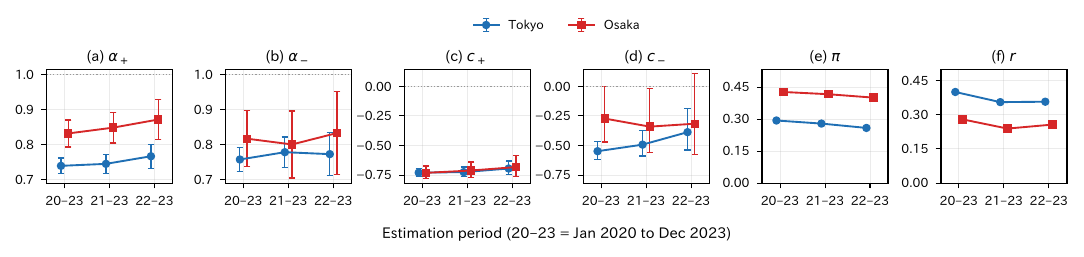}
\caption{Sensitivity of parameter estimates to the estimation period at 100\,m. All periods end in December 2023; starting dates are January 2020, 2021, and 2022. Error bars show 95\% intervals for the preference exponents and offsets.}
\label{fig:period_sensitivity}
\end{figure*}

\textbf{Sensitivity to the estimation period.}
At 100\,m, excluding 2020 or 2020--2021 preserves sublinear opening and closing estimates and negative opening offsets in both cities (Fig.~\ref{fig:period_sensitivity}). Closing estimates are less precise, especially in Osaka; $\pi$ and $r$ are slightly lower. This supports qualitative robustness but cannot separate coverage bias from temporal changes.

\begin{figure*}[t]
\centering
\includegraphics[width=0.97\textwidth,height=60mm,keepaspectratio]{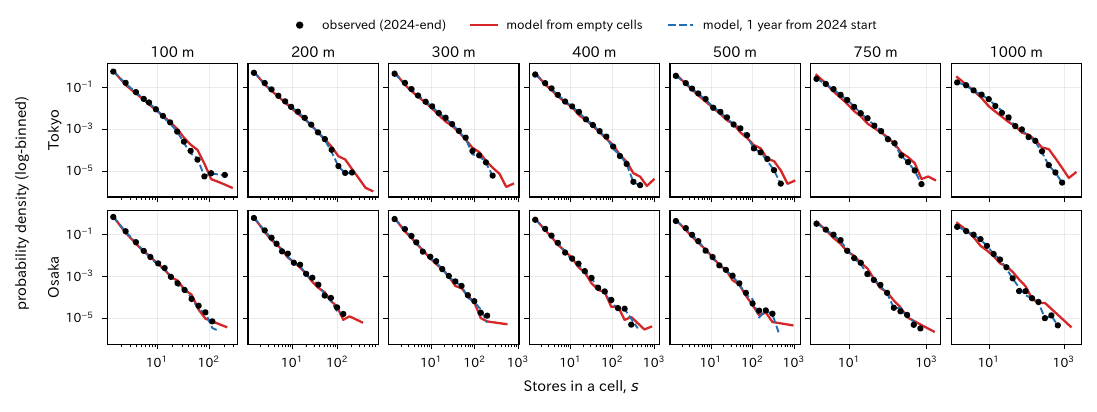}
\caption{Observed and simulated occupancy distributions (end of 2024; top: Tokyo, bottom: Osaka; 100--1000\,m from left). Black: observed; red: grown from empty cells; blue dashed: advanced from early 2024. Model curves are bin-wise medians over runs (five for red, 20 for blue).}
\label{fig:sim_dist}
\end{figure*}

\begin{figure*}[t]
\centering
\includegraphics[width=0.97\textwidth,height=65mm,keepaspectratio]{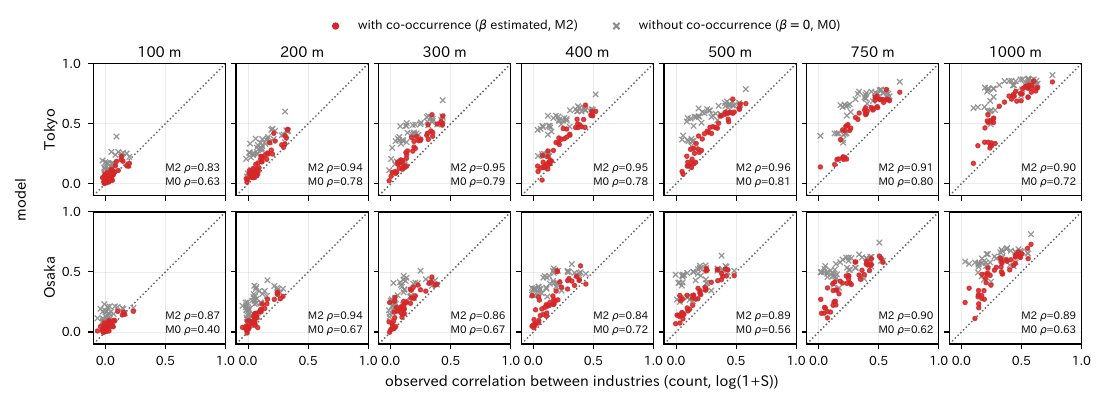}
\caption{Count correlations of the 45 industry pairs: observed (horizontal) vs.\ grown from empty cells (vertical). Red circles: M2 (with co-occurrence); gray crosses: M0 (without co-occurrence, $\beta=0$); dotted: identity. $\rho$ is the Pearson correlation over the pairs.}
\label{fig:sim_ind}
\end{figure*}

\section{Generative Simulation}\label{sec:simulation}
We run the model with the parameters estimated from 2020--2023 and compare with the end-2024 data, which were not used for estimation. Generation from empty cells tests the generative mechanism; generation from the early-2024 configuration is a one-year-ahead reproduction given the actual configuration and monthly event counts. The model has no spatial links between cells and does not generate geometric structure between neighboring cells. Because total store counts or event counts are given and occupied-cell counts depend strongly on $\pi$ and $r$, the validation is conditional and only partly independent of estimation.

\subsection{Generation Settings and Evaluation Metrics}
(i) For the occupancy distribution, we compare the log-binned density, the number of occupied cells, and the empirical CCDF slope (hereafter, the tail slope), reported as $-\tau$, of $P(S\ge s)$ on log--log axes for $s\ge3$. This common comparison range excludes one- and two-store cells; it is not a statistically estimated tail threshold. (ii) For co-location, we compute the correlations of $\log(1+S_{ij})$ between industries over occupied cells (count correlations) and those of the residuals after regression on $\log s_i$ (partial correlations), and measure the agreement between the model and the data over the 45 industry pairs by the Pearson correlation coefficient $\rho$ and the RMSE. We compare M2 (with co-occurrence, estimated $\beta$) and M0 (without co-occurrence, $\beta=0$); by Eq.~(\ref{eq:additivity}), $\beta$ does not affect occupancy.

\textbf{Generation from empty cells.} From an all-empty state, we mix openings and closings at $1:r$ until the total number of stores at the end of 2024; only the total numbers of cells and stores are given. Events amounting to 1\% of the total are processed per step, and we run five replicates; scalar metrics (Table~\ref{tab:comparison}) are averaged over runs.

\textbf{Generation from the early-2024 configuration.} From the actual configuration at the beginning of 2024, we apply the observed monthly numbers of openings and closings for 2024 (20 runs).

\subsection{Reproduction of Occupancy and Co-location}
\begin{table}[t]
\centering
\caption{Generation from empty cells (end of 2024; means of five runs). Count corr.: agreement ($\rho$ / RMSE) over 45 pairs. Partial: agreement of partial correlations. M2/M0: with/without co-occurrence.}\label{tab:comparison}
\setlength{\tabcolsep}{2.5pt}
\resizebox{\columnwidth}{!}{%
\begin{tabular}{rrrrrccc}
\toprule
 & \multicolumn{2}{c}{Occupied cells} & \multicolumn{2}{c}{Tail slope} & \multicolumn{2}{c}{Count corr.\ $\rho$ / RMSE} & Partial $\rho$ \\
$\ell$ [m] & Obs. & Model & Obs. & Model & M2 & M0 & M2 / M0 \\
\midrule
\multicolumn{8}{l}{Tokyo} \\
100 & 7{,}293 & 7{,}039 & $-1.82$ & $-1.81$ & 0.83 / 0.042 & 0.63 / 0.111 & 0.96 / 0.95 \\
200 & 4{,}676 & 4{,}226 & $-1.70$ & $-1.34$ & 0.94 / 0.080 & 0.78 / 0.188 & 0.90 / 0.86 \\
300 & 3{,}448 & 2{,}928 & $-1.49$ & $-1.13$ & 0.95 / 0.107 & 0.79 / 0.235 & 0.89 / 0.82 \\
400 & 2{,}709 & 2{,}096 & $-1.38$ & $-1.06$ & 0.95 / 0.110 & 0.78 / 0.268 & 0.89 / 0.75 \\
500 & 2{,}157 & 1{,}580 & $-1.29$ & $-0.96$ & 0.96 / 0.132 & 0.81 / 0.287 & 0.90 / 0.78 \\
750 & 1{,}396 & 837 & $-1.13$ & $-0.80$ & 0.91 / 0.179 & 0.80 / 0.300 & 0.90 / 0.78 \\
1000 & 936 & 382 & $-1.05$ & $-0.66$ & 0.90 / 0.236 & 0.72 / 0.387 & 0.89 / 0.73 \\
\midrule
\multicolumn{8}{l}{Osaka} \\
100 & 3{,}947 & 3{,}951 & $-1.70$ & $-1.57$ & 0.87 / 0.039 & 0.40 / 0.114 & 0.96 / 0.92 \\
200 & 2{,}872 & 2{,}716 & $-1.46$ & $-1.24$ & 0.94 / 0.064 & 0.67 / 0.193 & 0.94 / 0.88 \\
300 & 2{,}288 & 2{,}081 & $-1.37$ & $-1.11$ & 0.86 / 0.114 & 0.67 / 0.214 & 0.84 / 0.78 \\
400 & 1{,}943 & 1{,}639 & $-1.25$ & $-1.00$ & 0.84 / 0.131 & 0.72 / 0.276 & 0.82 / 0.79 \\
500 & 1{,}619 & 1{,}306 & $-1.20$ & $-0.99$ & 0.89 / 0.142 & 0.56 / 0.277 & 0.85 / 0.71 \\
750 & 1{,}110 & 725 & $-1.12$ & $-0.94$ & 0.90 / 0.175 & 0.62 / 0.321 & 0.88 / 0.61 \\
1000 & 794 & 446 & $-1.08$ & $-0.86$ & 0.89 / 0.193 & 0.63 / 0.324 & 0.87 / 0.63 \\
\bottomrule
\end{tabular}}
\end{table}

\textbf{Occupancy distribution.} At 100\,m, the model density almost coincides with the observations from $s=1$ to the tail (Fig.~\ref{fig:sim_dist}, Table~\ref{tab:comparison}). The number of occupied cells (Tokyo: 7{,}039 vs.\ 7{,}293 observed; Osaka: 3{,}951 vs.\ 3{,}947) and the tail slope (Tokyo: $-1.81$ vs.\ $-1.82$; Osaka: $-1.57$ vs.\ $-1.70$) agree, and the heavy tail emerges from an empty state. At larger cell sizes, the model has too few occupied cells (in Tokyo, 0.97 of the observation at 100\,m, 0.73 at 500\,m, and 0.41 at 1000\,m) and a flatter tail (Tokyo 500\,m: $-0.96$ vs.\ $-1.29$), although the flattening with cell size is shared (Tokyo: $-1.82$ to $-1.05$ observed, $-1.81$ to $-0.66$ model). Osaka shows smaller discrepancies in the same direction. Starting from the early-2024 configuration, the densities coincide with the observations at 100 and 500\,m in both cities.

\textbf{Inter-industry co-location.} The agreement of count correlations at 100\,m is $\rho=0.83$ (RMSE 0.042) in Tokyo and $\rho=0.87$ (0.039) in Osaka with co-occurrence, substantially exceeding $\rho=0.63$ and 0.40 without it (Fig.~\ref{fig:sim_ind}); M2 achieves better agreement at all cell sizes, although it has more parameters than M0. Without $\beta$, shared cell occupancy can still induce correlations. For partial correlations, both models agree well at 100\,m (Tokyo 0.96 vs.\ 0.95, Osaka 0.96 vs.\ 0.92), and M2's advantage tends to grow with cell size (Tokyo 500\,m: 0.90 vs.\ 0.78; Osaka 750\,m: 0.88 vs.\ 0.61). At large cell sizes, the model overestimates correlation levels and the RMSE grows, presumably because it concentrates stores in too few cells.

\section{Discussion}\label{sec:discussion}

The reproduction from an empty state at 100\,m suggests that the growth rule---new agglomerations are seeded at the rate $\pi$ and existing ones grow according to their size---can be a sufficient mechanism for the observed heavy tail. While the mean-field value ($\tau_{\rm th}=1.31$) did not match the observation, the simulation with $\alpha_+<1$ did. Indeed, changing only $\alpha_+$ from 0.74 to 0.8, 0.9, and 1.0 at 100\,m in Tokyo flattened the tail slope from $-1.84$ to $-1.54$, $-1.14$, and $-0.80$, comparable to multiplying $\pi$ by 1.5 or 0.5 ($-2.40$, $-1.38$). The tail is thus set both by the supply of new agglomerations and by the less-than-proportional growth of large ones. At large cell sizes, small $\pi$ and few empty cells lead the model to concentrate stores in too few cells, the regime where the finiteness of empty cells ignored in Section~\ref{sec:theory} matters. The co-location patterns in $\beta$ and $R_{jk}$ are largely common to both cities, while the cities differ mainly in $\pi$ and $r$.

\textbf{Limitations.} (i) Web coverage may vary by area, industry, and period; validation and correction using external, preferably longitudinal, POI data and official statistics remain necessary. (ii) Stores without observed events are omitted, and initial occupancy favors stores with subsequently observed closings, potentially biasing $\alpha_\pm$, $c_\pm$, $\pi$, and $\beta$. (iii) Missing closings may bias $\alpha_-$ and $r$ (Section~\ref{sec:theory}); the direction and magnitude of parameter bias are unknown. (iv) Generation from empty cells assumes constant historical $\pi$ and $r$. (v) Occupancy proxies location attractiveness rather than identifying causal reinforcement~\cite{bianconi2001}; explicit covariates and simpler location-choice baselines remain to be examined. (vi) Results depend on cell size and grid boundaries; distance-based co-location analysis remains future work. (vii) Distributional reproduction does not establish superiority over lognormal or other distributions, and uncertainty in empirical CCDF slopes remains unquantified.

\section{Conclusion}\label{sec:conclusion}
Using web-derived store opening and closing events in Tokyo and Osaka, we showed that the occupancy distribution has a heavy tail at several cell sizes and proposed a two-stage stochastic model of growth-type location choice and industry choice. Given the estimated parameters and the total number of stores, the model reproduced the end-2024 occupancy distribution, particularly at fine spatial resolutions, as well as inter-industry co-location. Its additive consistency allows nonlinear inter-industry interactions without changing the aggregate occupancy distribution.

Future work includes testing scale invariance through larger study areas and multiscale modeling, incorporating spatial links to represent spillovers and street-level agglomerations, validating data coverage, and extending the analysis to non-urban areas.

\FloatBarrier

\end{document}